# Asymmetric Electronic Band Alignment and Potentially Enhanced Thermoelectric Properties in Phase-Separated $Mg_2X$ (X=Si,Ge,Sn) Alloys

**Byungki Ryu,[1,7*] Samuel Foster,[2] Eun-Ae Choi,[3] Sungjin Park,[1] Jaywan Chung,[1] Johannes de Boor,[4,5] Pawel Ziolkowski,[4] Eckhard Müller,[4,6] Seung Zeon Han,[3] SuDong Park,[1] Neophytos Neophytou[2,&]**

[1] Energy Conversion Research Center, Electrical Materials Research Division, Korea Electrotechnology Research Institute (KERI), Changwon 51543, Republic of Korea.

[2] School of Engineering, University of Warwick, Coventry CV4 7AL, UK

[3] Extreme Materials Research Institute, Korea Institute of Materials Science (KIMS), Changwon 51508, Republic of Korea.

[4] German Aerospace Center (DLR) - Institute for Frontier Materials on Earth and in Space, 51170 Cologne, Germany

[5] Faculty of Engineering, Institute of Technology for Nanostructures (NST) and CENIDE, University of Duisburg-Essen, Building BA, Bismarckstr. 81, D–47057 Duisburg, Germany

[6] Institute of Inorganic and Analytical Chemistry, Justus Liebig University, Giessen, 35392 Giessen, Germany

[7] Electric Energy and Materials Engineering, School of KERI, University of Science and Technology, Changwon 51543, Republic of Korea

Correspondence: * byungkiryu@keri.re.kr (BR), & n.neophytou@warwick.ac.uk (NN)

## Abstract

The $Mg_2X$ (X=Si, Ge, Sn) based alloy is an eco-friendly thermoelectric material for mid-temperature applications. The $Mg_2Si_{1-x}Sn_x$ and $Mg_2Ge_{1-x}Sn_x$ alloys can be phase-separated into Si(Ge)- and Sn-rich phases during material synthesis, leading to a nanocomposite with locally varying electronic band structure. First-principles calculations reveal that the valence band offset is eight-times larger than the conduction band offset at the interface between Si- and Sn-rich phases for x = 0.6, showing type-I and asymmetric band alignment (0.092 eV versus 0.013 eV). Using Boltzmann transport theory and thermionic emission calculations, we show that the large valence band energy discontinuity could allow for energy filtering effects to take place that can potentially increase the power factor substantially in the *p*-type material system if designed appropriately.

# I. INTRODUCTION

$Mg_2Si$-based thermoelectric alloys ($Mg_2X$, X=Si, Ge, Sn, and their alloys), consisting of non-toxic and earth-abundant elements, have garnered much attention owing to their high n-type $ZT$ in the mid-temperature range (600–800 K).[1–3] From thermodynamic calculations and some of the experimental data, there exists a large composition range where $Mg_2Si$–$Mg_2Sn$ (or $Mg_2Ge$–$Mg_2Sn$) alloys are immiscible, with the extent of this miscibility gap decreasing with increasing temperature due to entropy.[4–7] On the other hand, single phase materials have repeatedly been synthesized within the predicted miscibility gap.[8–10] It has been argued that coherent interfaces formed during synthesis stabilize the material against unmixing and that breaking this coherence can be an experimentally feasible path to trigger and influence the alloy nanostructuring.[6,11]As a result, by adjusting processing conditions (e.g. temperature) these alloys can be transformed into Si-rich $Mg_2Si_{1-x}Sn_x$ and Sn-rich $Mg_2Si_{1-x}Sn_x$ phases due to the miscibility gap, forming an interface-rich microstructure, which can be an effective method to generate a complex nanostructure for phonon scattering, leading to a reduced thermal conductivity.[12]

In general, interfaces also control charge transport depending on the band offset and charge carrier energy (position of the Fermi level $\eta_F$),[13] which often proves beneficial for improving the performance of various electrical and optical devices, and solar cells.[14–16] Similarly, in thermoelectric materials, although the electrical conductivity can suffer, the interfacial energy barriers at the phase boundaries could play a critical role in thermoelectric transport via energy filtering[17–22] and bipolar transport suppression.[23–26] Despite the current large interest in the bulk $Mg_2X$ alloy systems, and some early experimental work probing for energy filtering,[27] there are only limited reports on the band alignment of the $Mg_2Si$–$Mg_2Sn$ alloy system, and how that can prove beneficial to its thermoelectric properties. Hence, the role of interfaces on the electronic transport in this system is not well understood at present.

In this paper, we report on theoretical results for the *electronic band energy alignment* between phase-separated $Mg_2Si_{1-x}Sn_x$ alloys and its potential effect on thermoelectric transport (power factor) across the phase boundaries. From first-principles calculations, we find an *asymmetric* band offset with type-I band alignment between Si- and Sn-rich $Mg_2$(Si,Sn) phases. Using Boltzmann transport theory and thermionic emission, we show that the phase-separated alloy system can provide energy filtering and enhanced power factors (PFs) for the *p*-type system due to the large valence band energy discontinuity.

# II. CALCULATION METHOD

We performed first-principles density-functional theory (DFT)[28,29] calculations for atomic and electronic structure of $Mg_2X$ using projector-augmented wave pseudopotentials,[30,31] and generalized-gradient approximation (GGA),[32] as implemented in the Vienna Ab initio Simulation Package (VASP) code.[33,34] We fully optimized atomic structure within GGA calculations structures, while the lattice parameters of the solid-solution supercells were interpolated using those of the corresponding binary phases. The mixing energies of $Mg_2$(Si,Sn) solid solutions were calculated compared to ideal binary phases of $Mg_2Si$ and $Mg_2Sn$ and the mixability gap was predicted with inclusion of an ideal configuration entropy. To obtain reliable electronic structures, we also carried out many-body perturbation theory (MBPT) non-self-consistent $G_0W_0$ approximation calculations for bulk $Mg_2Si$, $Mg_2Ge$, and $Mg_2Sn$ on top of the hybrid-DFT.[35] For this, we employed the HSE06 functional with the exchange mixing parameter of 35% and screening parameter of 0.2 Å$^{-1}$,[35,36] hereafter referred to as HSE-35. The details for band structure calculations can be found in a previous report.[36] Note that spin-orbit interaction (SOI) is included for all electronic structure calculations.

For the band alignment of $Mg_2X$ alloys and related materials, the reference potential method[37–41] was applied on the ideal 16-layer (110) $Mg_2X$ surface structures. We extracted the reference potential of $Mg_2X$ surface structures with respect to the vacuum energy. For simplicity, the strain on bulk and interfaces is neglected. By aligning the $G_0W_0$ corrected bulk band edge energies on the DFT surface reference potential,[38,41] the band edge energies are aligned with the zero vacuum level. For the ternary alloys, one valence band and two conduction band minima energies were linearly interpolated with composition to predict the composition-dependent band structure. Around the conduction band convergence composition, the two conduction bands cross and then separate again. Considering the character of each band, we tracked the evolution of each band edge individually and performed interpolations accordingly.

# III. RESULTS AND DISCUSSION

**Figure 1(a)** and **(b)** show the atomic structures of the 96-atom supercells for binary $Mg_2Si$ and solid solution $Mg_2Si_{0.59}Sn_{0.41}$. The ideal $Mg_2X$ compounds has an antifluorite structure, where the anion X atoms are positioned at the FCC regular sites and the cation Mg atoms are positioned at the tetrahedral interstitial sites. In this configuration, each X anion atom is surrounded by eight Mg atoms. We used the optimized lattice parameters of 6.348 Å, 6.410 Å, and 6.801 Å for $Mg_2Si$, $Mg_2Ge$, and $Mg_2Sn$, as predicted by the previous study.[36] We modelled the solid solution of ternary $Mg_2(Si_{1-x}Sn_x)$ by assuming a random substitution of Si and Sn atoms at the X sites, with the lattice parameter linearly interpolated between those of $Mg_2Si$ and $Mg_2Sn$. After structural relaxation, the positions of Mg, Si, and Sn elements were displaced from the ideal positions.

**Figure 1(c)** quantifies the structural disorder in solid solution models using the root mean square (RMS) displacement, defined as the atomic position displacement from their ideal lattice sites after relaxation. On average, the RMS displacement was smaller than 0.06 Å. Mg exhibits the largest displacement, followed by Si. Sn atoms show relatively small displacement for composition x < 0.6, but is becomes comparable to that of Si when x > 0.6.

**Figure 2** shows the DFT-based predictions of the thermodynamic stability and phase diagram of $Mg_2(Si,Sn)$. The $Mg_2Si$ and $Mg_2Sn$ alloys are immiscible at 0 K, but their alloy becomes miscible at elevated temperatures. A randomly generated supercell structure of $Mg_2(Si,Sn)$ exhibits a positive mixing energy ($E_{mix}$), consistent to the immiscibility reported in the phase diagram and various literature reports.[4,5,42] However, when sufficient thermal energy is available, a solid solution can form. To quantitatively analyze the thermodynamic within the DFT framework, we incorporate an ideal configuration entropy ($S_{mix}$), assuming an ideal solid solution. The temperature-dependent mixing free energy is calculated as:

$$\mathrm{E_{mix}} = \mathrm{E[Mg_2Si_{1-x}Sn_x]} - \{(1-\mathrm{x}) \cdot \mathrm{E[Mg_2Si]} + \mathrm{x} \cdot \mathrm{E[Mg_2Si]}\}\,, \quad \textbf{(1)}$$

$$\frac{\mathrm{S_{mix}}(x)}{\mathrm{k_B}} = \Sigma(-p_i \ln(p_i)) \approx -x\ln(x) - (1-x)\ln(1-x)\,, \quad \textbf{(2)}$$

$$\mathrm{F_{mix}} = \mathrm{E_{mix}} - \mathrm{TS_{mix}}\,. \quad \textbf{(3)}$$

**Figure 2(a)** shows the calculated mixing energy $E_{mix}$ for $Mg_2(Si_{1-x}Sn_x)$ alloys obtained from DFT calculations with respect to two end members, $Mg_2Si$ and $Mg_2Sn$. The $E_{mix}$ exceeds 40 meV per formula unit at intermediate compositions, indicating strong immiscibility at low temperatures. $E_{mix}$ and $F_{mix}$ were fitted to a fourth-degree degree polynomial. **Figure 2(b)** shows the temperature-dependent mixing free energy. As temperature increases, the entropic contribution lowers the free energy, thereby stabilizing the solid solution state.

At 700 K, shown in **Figure 2(c)**, the $F_{mix}$ becomes negative near the Si-rich (x < 0.1) and Sn-rich (x > 0.9) ends, suggesting that phase coexistence of two different composition is thermodynamically favorable, consistent with a previous work, but with different compositions.[7] The double well shape of the free energy curve indicates the possibility of phase separation with intermediate compositions. The binodal points, marked by filled circles, represent the boundaries of two-phase region and are determined from the common tangent construction:

$$\left.\frac{\partial F}{\partial x}\right|_{x=x_{1b}} = \left.\frac{\partial F}{\partial x}\right|_{x=x_{2b}}. \quad (4)$$

The derivative of the free energy with respect to composition corresponds to the chemical potential. Thus, the binodal points represent compositions at which two phases coexist in equilibrium with equal chemical potentials. The spinodal points, marked by open stars, corresponds to inflection points of the free energy curve, where the second derivative vanishes:

$$\left.\frac{\partial^2 F}{\partial x^2}\right|_{x_{1s}} = \left.\frac{\partial^2 F}{\partial x^2}\right|_{x_{2s}} = 0. \quad (5)$$

When the curvature of the free energy (the second derivative) is positive, the system is locally stable against small perturbations in composition. However, inside the spinodal region, where the curvature becomes negative, the system is unstable with respect to such compositional fluctuations, and spinodal decomposition occurs. **Figure 2(d)** summarizes the temperature-dependent phase diagram with binodal and spinodal boundaries. The binodal gap is wider than the spinodal gap. For instance, at 1200 K, a two-phase region exists between Si-rich $Mg_2Si_{0.6}Sn_{0.4}$ and Sn-rich $Mg_2Si_{0.4}Sn_{0.6}$. As temperature decreases, the miscibility gap widens. Below 400 K, the two binaries are nearly completely immiscible. Although the spinodal decomposition gap is relatively large, the second derivative of free energy remains nearly constant over a wide composition range. This implies that small compositional fluctuations within the spinodal region do not significantly lower the free energy, supporting the likelihood of a de-mixed two-phase state in Mg-Si-Sn alloys.

**Table 1** summarizes the band gap values for the $Mg_2Si$, $Mg_2Ge$, and $Mg_2Sn$ obtained from various calculation methods and experimental measurement techniques. Within DFT-PBE, these binary compounds have band gaps smaller than 0.2 eV, indicating a narrow band gap semiconductor. When more advanced calculation methods such as hybrid-DFT or GW many body approaches are employed,[36,43,44] the calculated band gaps show much better agreement with experimental values.[45–48]

**Figures 3(a)** and **3(b)** show the calculated electronic band alignment for the $Mg_2Si$, $Mg_2Ge$, and $Mg_2Sn$ binaries with respect to the vacuum energy levels. In DFT calculations, the band alignment is not reliable due to the well-known band gap underestimation issue. For $Mg_2Sn$, the DFT band gap is even negative, inconsistent with previous observations.[43,45] To overcome this, we calculated the band alignment using MBPT $G_0W_0$ calculations, which provide corrected band gaps for $Mg_2X$. Finally, a type-I band alignment, where both the conduction and valence band edges of one component lie within the band gap of the other, is obtained. The alignment is also asymmetric due to the different offsets at the conduction and valence band edges, as shown in **Figure 3(b)**. The conduction band offset (CBO) is smaller than the valence band offset (VBO). In the MBPT $G_0W_0$ calculations, the corrected CBM energy of $Mg_2Si$ ($Mg_2Ge$) was calculated to be 0.13 eV (0.10 eV), and the VBM of $Mg_2Si$ ($Mg_2Ge$) was calculated to be −0.46 eV (−0.42 eV) with respect to the CBM (VBM) of $Mg_2Sn$, respectively.

**Figure 3(c)** shows the electronic band alignment for $Mg_2Si_{1-x}Sn_x$ ternaries and related materials. For the Si- and Sn-rich phases of $Mg_2(Si,Sn)$, we consider the $Mg_2Si_{0.6}Sn_{0.4}$ and $Mg_2Si_{0.4}Sn_{0.6}$ compositions. These compositions correspond to the binodal points at 1200 K, where phase separation into Si-rich and Sn-rich domains is thermodynamically favorable according to the calculated phase diagram. It is worth noting that we interpolated the conduction band energies of the $X_1$ and $X_3$ states in binary $Mg_2Si$ and $Mg_2Sn$, and obtained the CBM energy of the solid-solution states by considering the band crossing between $X_1$ and $X_3$ in $Mg_2(Si,Sn)$ alloys. From the MBPT $G_0W_0$ calculations, the CBO at the interface between Sn-rich $Mg_2Si_{0.4}Sn_{0.6}$ and Si-rich $Mg_2Si_{0.6}Sn_{0.4}$ phases is relatively small (0.013 eV), whereas the VBO is larger (0.092 eV). We also compare the relative band alignment between $Mg_2X$ and related elemental materials: $Mg_2Si$, Si-rich $Mg_2$(Si,Sn), Sn-rich $Mg_2$(Si,Sn), $Mg_2Sn$, metallic Mg, semiconducting Si and Ge in diamond structure, and metallic Sn. For the metallic elements, we used their experimental work functions: 3.68 eV for Mg and 4.42 eV for Sn.[41] For the semiconducting element phases, we use the reported band alignment values: 3.7 eV for n-type Si and 3.5 eV for n-type Ge.[49] Interestingly, we also find an asymmetric behavior between $Mg_2X$ and X elements. Other than for Sn, the CBMs and Fermi level of metals are more aligned, compared to their VBMs.

In $Mg_2$(Si,Sn) alloys, the conduction band convergence of the X1 and X3 CBM states is known to be the origin of the high power factor in the n-type system.[8,36] Furthermore, the significant change in the valence band edge energies between the $Mg_2X$ based binary and ternary alloys could suggest an additional mechanism for enhanced thermoelectric performance in the phase-separated *p*-type $Mg_2Si$-$Mg_2Sn$ alloys. The band offset magnitudes of Si- and Sn-rich phases are large enough to be able to act as barriers for carrier energy filtering in the case of *p*-type transport.[22]

To provide a first order upper level estimate of the power factor improvement that filtering can allow in the phase-separated $Mg_2Si_{0.4}Sn_{0.6}$/$Mg_2Si_{0.6}Sn_{0.4}$ alloy system compared to the single phase $Mg_2Si_{0.4}Sn_{0.6}$ (the one with the lowest bandgap, which will act as the base material reference), we performed transport calculations based on the Boltzmann Transport Equation (BTE). To simplify the transport treatment, we assume that it is dominated by elastic, isotropic scattering mechanisms (like acoustic phonon scattering,[10,50]) which allows the use of simplified expressions for the scattering rates. In BTE, the transport coefficients are given by:

$$\sigma = R^{(0)} \qquad \textbf{(6)}$$

$$S = \frac{k_B}{q_0}\frac{R^{(1)}}{R^{(0)}} \quad (7)$$

$$\kappa_e = \frac{k_B^2 T}{q_0^2}\left[R^{(2)} - \frac{\left[R^{(1)}\right]^2}{R^{(0)}}\right] \quad (8)$$

where

$$R^{(\alpha)} = q_0^2 \int_{E_0}^{\infty} dE \left(-\frac{\partial f_0}{\partial E}\right) G(E) \left(\frac{E - E_F}{k_B T}\right)^{\alpha} \quad (9)$$

with the energy-dependent transport distribution function defined as:

$$G(E) = \tau_s(E)\, v^2(E) g(E)\,. \quad (10)$$

Above $\tau_s(E)$ is the relaxation time, $v(E)$ is the carrier velocity, and $g(E)$ is the density of states of the material. For simplicity, we assume parabolic effective mass bands with a global $m^* = 0.5m_0$ in all instances. We also assumed a constant (in energy) mean-free-path, MFP, i.e. $mfp = \tau_s(E)\, v(E) = 10\, nm$, which infers acoustic phonon scattering limited transport (without loss in generality here). Here, the *mfp* value we use is in the order of what encountered for phonon-limited transport is typical semiconductors. We do not have specific accurate information for the *mfp* of the alloys under investigation, not we include details if ionized impurity scattering, boundary scattering, or alloy scattering. Thus, our results should only provide qualitative estimates, rather than quantitative predictions, but still they will point towards an upper limit PF improvement estimation (at first order).

In order to cover the different length scales that the two different phases can have, and how these will affect the TE performance, we assume two situations: i) the individual phase domains are large regions, much larger compared to the characteristic lengths of electron momentum and energy relaxation. In this case the charge carriers are in equilibrium in each region, and a series resistance model can be employed. At first order, the electronic conductivity and Seebeck coefficient in a two-phase material are given by:

$$\frac{v_{tot}}{\sigma_{\text{tot}}} = \frac{V_1}{\sigma_1} + \frac{V_2}{\sigma_2} \tag{11}$$

$$S_{\text{tot}} = \frac{S_1 V_1 + S_2 V_2}{V_1 + V_2} \tag{12}$$

where the total is given by combinations of the individual phase components, weighted by the volume of the different phases (which in this case we assume for them to be equal).

In the second scenario we assume that the individual phase regions are small, to enable non-equilibrium transport and enable an effective energy filtering design mechanism as described in our previous works.[51,52] In this case we assume that the barrier regions are narrow enough to enable thermionic emission without the carrier energies relaxing on the barriers, while the wells are of the order of the energy relaxation length of the carriers, such that they do not relax their energy in the wells. In this case, high energy (hot), high velocity and high mobility carriers are injected from the barriers over the wells, while the low energy, cold carriers are blocked. The latter increases the Seebeck coefficient, which is retained high both in the barriers and wells, as carriers do not relax their energy in the wells. The former allows for retained conductivity (despite the presence of the barriers), because the thermionically emitted carriers are of high velocity and mobility. In the presence of a potential barrier, as is the case at the interfaces between dissimilar materials, in which case potential barriers of height $V_B$ are formed for carriers, it is customary to impose thermionic emission above the barriers as, under which the conductivity is given by:

$$\sigma(E) = 0 \quad \text{for } E \leq V_B, \tag{13}$$

$$\sigma(E) = \sigma_0 T_r(E) \quad \text{for } E > V_B, \tag{14}$$

where $\sigma(E)$ and $\sigma_0(E)$ are the carrier band energy dependent conductivity with and without potential barriers, and $T_r(E)$ is the transmission of charge energy state $E$ (here for simplicity we assume $T_r(E) = 1)$ . In this case, transport over the barriers provide energy filtering, which increases the Seebeck coefficient composite.

Thus, we examine two cases: i) the relative performance of the two-phase system if the phases are large enough such that transport in each phase is resistive, and ii) the relative performance of the *p*-type system in the presence of energy filtering. This is done in a single way by shifting the Fermi level from the conduction band to the valence band and computing the transport properties. Here the transport properties of the electron and hole carriers are computed separately and then combined as described in Ref.[26]

**Figure 4** shows the TE coefficients (conductivity $\sigma$, Seebeck coefficient $S$, and power factor $\sigma S^2$) versus the position of the Fermi level $\eta_F$ for three different cases: i) the single phase $Mg_2Si_{0.4}Sn_{0.6}$ system which will act as a reference (black lines), ii) the $Mg_2Si_{0.4}Sn_{0.6}$–$Mg_2Si_{0.6}Sn_{0.4}$ two-phase system under resistive transport conditions, assuming that the individual phases are large enough such that charge carriers relax their energies in the bands of the phase and are in near-equilibrium conditions (blue lines), and iii) the $Mg_2Si_{0.4}Sn_{0.6}$–$Mg_2Si_{0.6}Sn_{0.4}$ two-phase system the larger-gap phase $Mg_2Si_{0.6}Sn_{0.4}$ acting as the filtering barrier for holes – a 0.092 eV barrier is formed in the VB and 0.013 eV in the CB (red lines) – in this case we assume that the phases are short enough such that charge carriers do not fully relax in the wells or the barriers, having non-equilibrium transport features, such that the band discontinuities can act effectively as filtering barriers. This can be typically between 30 nm up to 100 nm in common semiconductors. All simulations here are performed at room temperature of $T$ = 300 K. We then observed whether the phase-separated alloy system has the ability to improve the PF, in addition to the expected reduction in its thermal conductivity (not examined here).

The conductivity of the three material cases examine has a very similar trend, with the only difference being the onset of the conductivity in the valence band, as modified by the band edge discontinuity (see **Fig. 4(a)**). The small differences in the conduction band make the systems behave identically. In the valence band, a sharper slope is acquired by the filtering system (red line), a typical feature of *thermionic emission*.[51,53] In **Fig. 4(b)**, the Seebeck coefficients of the two-phase systems are shifted towards the left in the valence band, a signature of increased resistance for the resistive system, and energy filtering for the filtering system. The increase in the Seebeck coefficient in regions of steep conductivity slope (red line), provides a large power factor increase as observed in **Fig. 4(c)**. In this case the PF can double, but this is an upper limit of what can be achieved. In practice the well/barrier geometry needs to be properly designed in order to retain this value,[22,26] but in this work we do not go into those details, we only want to demonstrate that the system at hand ($Mg_2Sn/Mg_2Si$-based phase-separated alloy system) could have potential for higher *p*-type PFs once designed appropriately to take advantage of energy filtering. Also note that the absolute values for the conductivity and the PF would be overestimated, since we have not included the effect of ionized impurity scattering when we shift the Fermi level into degenerate conditions, and we omit other scattering mechanisms such as optical and polar phonon scattering. Nevertheless, these will be common for all systems, such as the relative performance increase would still be a valid indication of the system behavior.

On the other hand, the resistive system (blue line) has lower PF performance for holes due to the reduction in the conductivity of the higher bandgap phase, and the fact that the regions are assumed large enough such that thermionic emission is absent, and the material reduced to a series resistance system. Here we are mostly interested in the *p*-type system, however, the large band discontinuity in the VB and its reduced transport under resistive conditions, could also lead to reduced electronic thermal conductivity and bipolar transport at high operating temperatures when the *n*-type polarity is considered, beyond the reduction in the lattice thermal conductivity, both of which are beneficial for larger *ZT* (although our calculations for this particular system indicate that the benefits are only marginal due to the relatively large bandgap of 0.61 eV).

Since electronic conduction in *p*-type system is degraded by the $Mg_2Si$ barrier, thus, for such phase-separated samples, this asymmetric band alignment can be responsible for the experimentally observed low mobility, in addition to the larger valence band mass in *p*-type alloys.[54,55] In that case, it is suggested that the material is doped to highly degenerate *p*-type conditions, such that the Fermi level is pushed into the valence bands closer to the barrier heights formed at the interfaces, and in that case power factor improvements are possible by taking advantage of filtering effects and high carrier velocities, a design strategy described in detail in Refs.[53,56] At the same time, our results also suggest that the asymmetric transport behavior plays a crucial role in metallic or semi-metallic thermoelectrics, consistent with earlier reports. If the system is heavily degenerately doped (deep Fermi level inside a conduction or valence band), the system will be highly metallic, but will still show high power factor, an unconventional feature compared semiconducting thermoelectrics. As explained in Ref.[57] transport asymmetry is what dictates larger Seebeck coefficients, and these can be achieved even under metallic conditions.[57] Meanwhile, *p*-type doping is challenging for this alloy.[58] Nevertheless, if such *p*-doping strategies can be successfully implemented such as using Li impurity doping,[59,60] they may enable enhanced power factor through energy filtering and high carrier velocities.

# IV. CONCLUSION

In summary, we report on the asymmetric electronic band alignments in $Mg_2X$ (X=Si, Ge, Sn) and their alloys. Owing to the immiscible nature, the $Mg_2$(Si,Sn) based alloys can be decomposed into Si- and Sn-rich phases at lower temperature range: $Mg_2Si_{0.6}Sn_{0.4}$–$Mg_2Si_{0.4}Sn_{0.6}$. In *p*-type alloys, the large valence band offset, which can act as a barrier, could be designed to provide energy filtering capabilities that could ideally provide large power factor improvements.

# ACKNOWLEDGMENTS

This work was supported by the Energy Efficiency & Resource Core Technology Program of the Korea Institute of Energy Technology Evaluation and Planning (KETEP) granted from the Ministry of Trade, Industry & Energy (MOTIE), Republic of Korea (Grant No, 2021202080023D). The work was also [58]through the National Research Council of Science and Technology (NST) funded by the Ministry of Science and ICT (MSIT) of the Republic of Korea (No. 25A01013). BR, EAC, and SZH is supported by the National Research Foundation of Korea (NRF) grant funded by the Korea Government (MSIP) (2022M3C1C8093916). JdB is partially funded by the Deutsche Forschungsgemeinschaft (DFG, German Research Foundation)-project number 520487260. NN received funding from the European Research Council (ERC) under the European Union's Horizon 2020 Research and Innovation Programme (Grant Agreement No. 678763) and from the U.K. Research and Innovation fund (Project Reference No. EP/X02346X/1).

# AUTHOR DECLARATION

## Conflict of Interest

The authors have no conflicts to disclose.

## Author Contributions

**Byungki Ryu:** Conceptualization; Methodology; Software; Validation; Formal analysis; Investigation; Resources; Data Curation; Writing - Original Draft; Writing - Review & Editing; Visualization; Project administration; Funding acquisition. **Samuel Foster:** Investigation. **Eun-Ae Choi:** Methodology; Validation; Formal analysis; Investigation; Writing - Review & Editing. **Sungjin Park:** Investigation. **Jaywan Chung:** Software; Investigation; Project administration; Funding acquisition. **Johannes de Boor:** Conceptualization; Writing - Review & Editing. **Pawel Ziolkowski:** Writing - Review & Editing. **Eckhard Müller:** Writing - Review & Editing. **Seung Zeon Han:** Investigation; Writing - Review & Editing. **SuDong Park:** Conceptualization; Supervision; Project administration; Funding acquisition. **Neophytos Neophytou:** Conceptualization; Methodology; Software; Validation; Formal analysis; Investigation; Writing - Original Draft; Writing - Review & Editing; Visualization; Supervision; Project administration; Funding acquisition.

# DATA AVAILABILITY

The data that support the findings of this study are available from the corresponding authors upon reasonable request: BR for band alignments and NN for transport calculations.

**Table 1.** Summary of band gap values for $Mg_2Si$, $Mg_2Ge$, and $Mg_2Sn$ compounds obtained from different theoretical approaches (DFT, hybrid functionals, GW) and experimental techniques (optical absorption and transport measurements).

| | **Method, Ref.** | **Band gap (eV)** | | |
|---|---|---|---|---|
| | | **$Mg_2Si$** | **$Mg_2Ge$** | **$Mg_2Sn$** |
| **Theory** | PBE with SOI (calculated lattice), This work, Ryu (2019) **[Ref. 36]** | 0.191 | 0.090 | -0.346 |
| | HSE-25 (calculated lattice), This work, Ryu (2019) **[Ref. 36]** | 0.568 | 0.523 | 0.146 |
| | HSE-25 (experimental lattice) Ayachi (2021) **[Ref. 44]** | 0.570 | - | 0.145 |
| | HSE-35 with SOI (calculated lattice), This work, Ryu (2019) **[Ref. 36]** | 0.715 | 0.617 | -0.030 |
| | $G_0W_0$ on top of HSE-25 with SOC (calculated lattice) Ryu (2019) **[Ref. 36]** | 0.763 | 0.690 | 0.077 |
| | $G_0W_0$ on top of HSE-30 with SOC (calculated lattice) Ryu (2019) **[Ref. 36]** | 0.798 | 0.724 | 0.177 |
| | $G_0W_0$ on top of HSE-35 with SOC (calculated lattice) This work, Ryu (2019) **[Ref. 36]** | 0.835 | 0.759 | 0.244 |
| | GW+SO, Shi and Kioupakis (2018) **[Ref. 43]** | 0.728 | 0.555 | 0.142 |
| **Experiment** | Optical, absorption coeff., Stella (1967) **[Ref. 45]** | 0.623 | 0.595 | - |
| | Logarithmic plot of electrical resistivity, Aizawa (2005) **[Ref. 46]** | 0.71 | 0.54 | - |
| | Infrared absorption, T-dep, Lott and Lynch (1966) **[Ref. 47]** | - | 0.57 | - |
| | Infrared absorption, T-dep, Lipson (1964) **[Ref. 48]** | - | - | 0.22-0.30 |

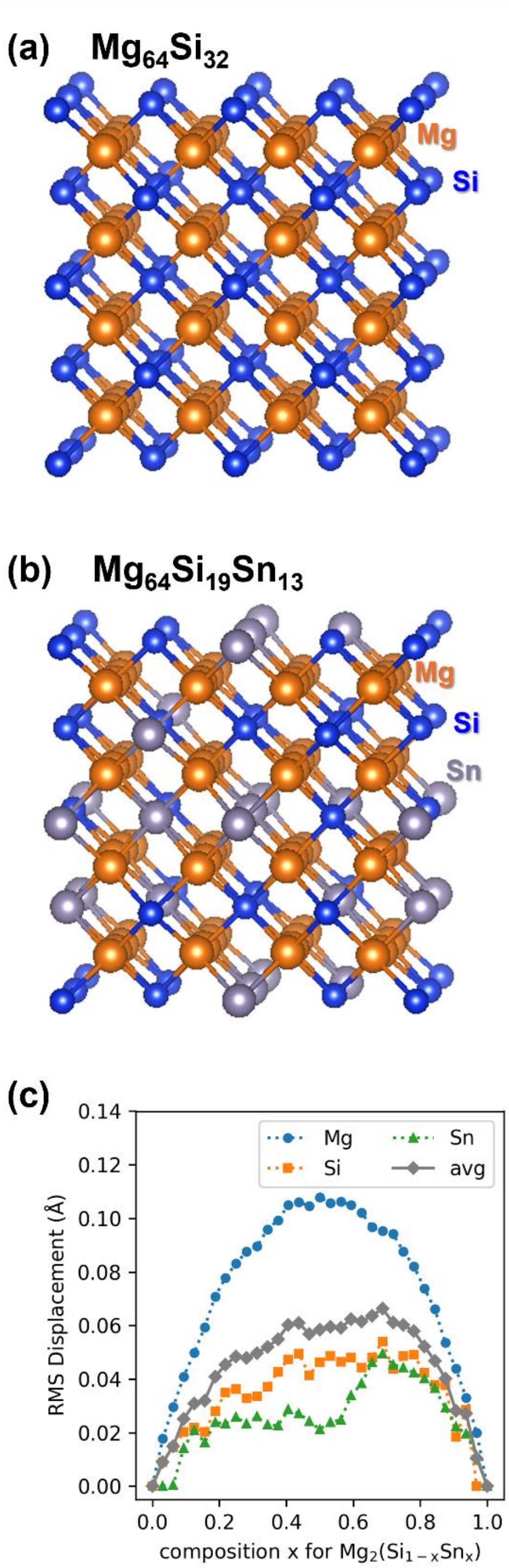

**Fig. 1 Atomic structure and displacement behavior of the $Mg_2Si_{1-x}Sn_x$ solid solution model.** Supercell models of (a) $Mg_{64}Si_{32}$ representing for binary $Mg_2Si$ and (b) $Mg_{64}Si_{19}Sn_{13}$ representing for $Mg_2Si_{0.59}Sn_{0.41}$ solid solution. (c) Calculated root mean-square displacement of Mg, Si, and Sn atoms versus composition x in $Mg_2Si_{1-x}Sn_x$, along with the average displacement.

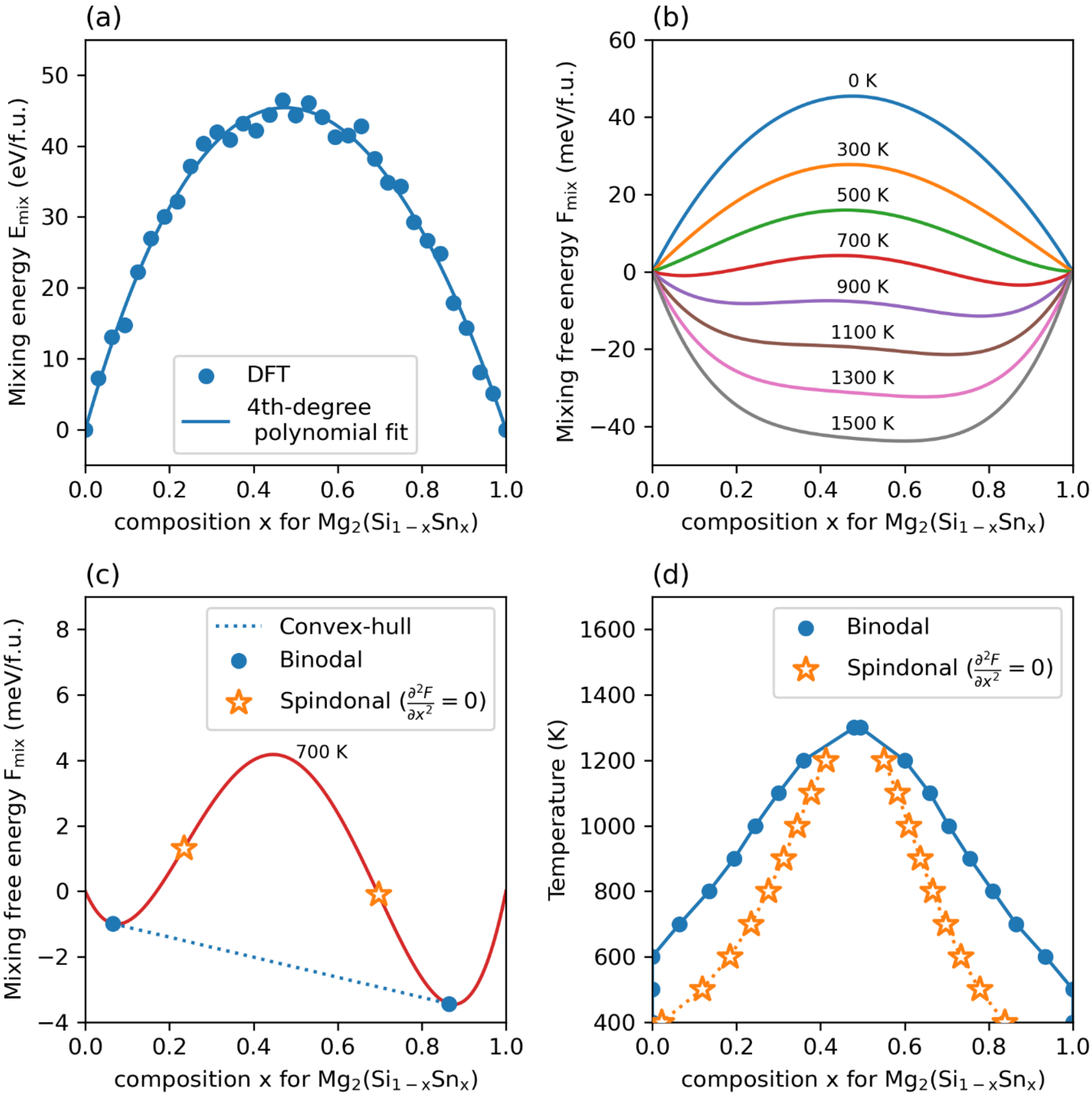

**Fig. 2 DFT-based prediction of the thermodynamic stability and phase diagram of $Mg_2Si_{1-x}Sn_x$ alloy systems.** (a) $E_{mix}$ and (b) temperature-dependent $F_{mix}$ curves with a $4^{th}$-degree polynomial fit. (c) Free energy at 700 K with binodal points from the convex hull and spinodal points from inflection points. (d) Calculated phase diagram showing binodal and spinodal boundaries over temperature.

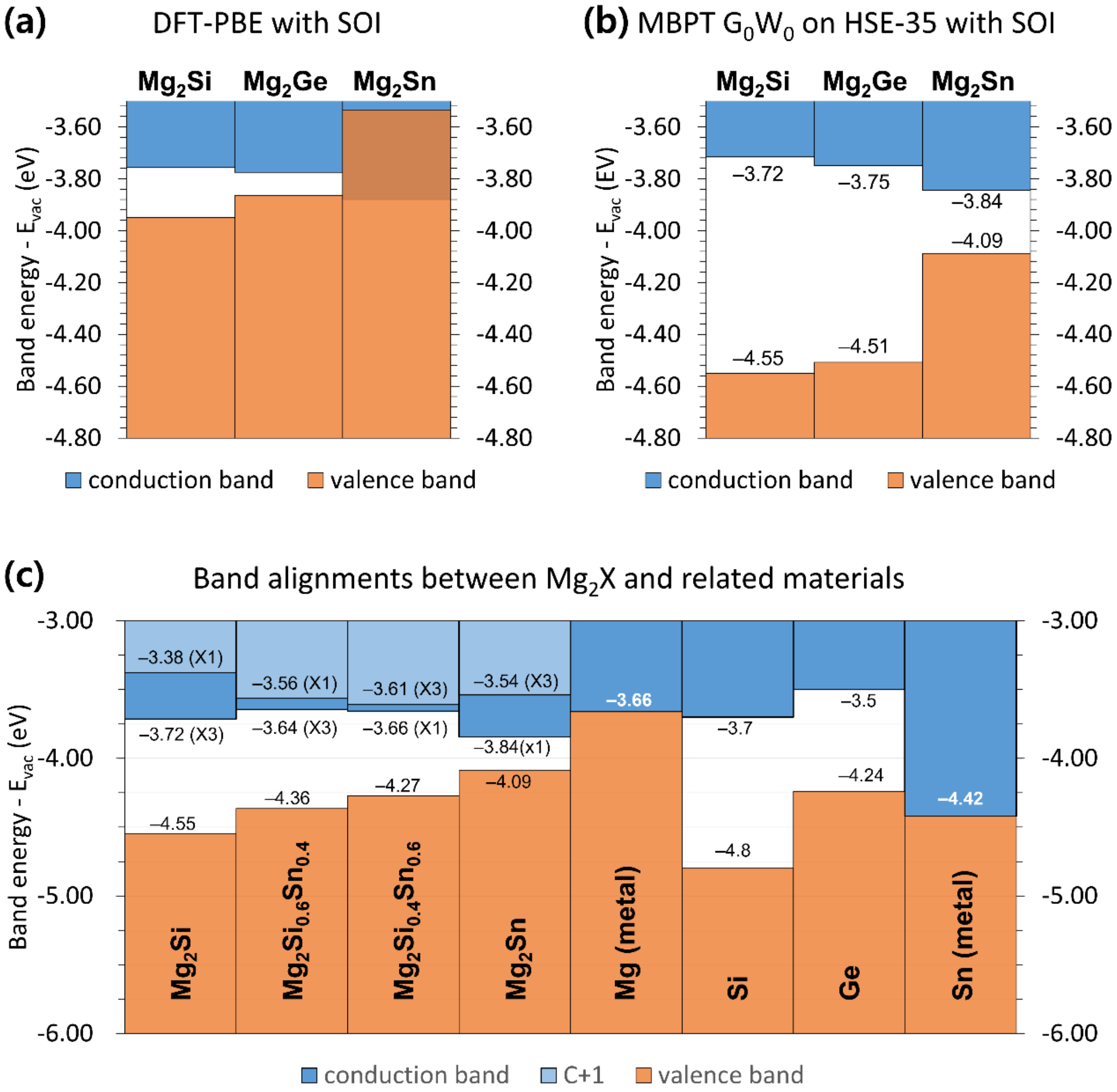

**Fig. 3** **Electronic Band alignment for $Mg_2X$ (X=Si, Ge, Sn)** obtained using: (a) the DFT-PBE calculations with SOI and (b) the MBPT $G_0W_0$ calculations on top of HSE35 with SOI calculations. In (c), the band edge positions of $Mg_2X$-related ternary alloys are shown and compared with the related elements. For alignment, the vacuum energy level is set to zero.

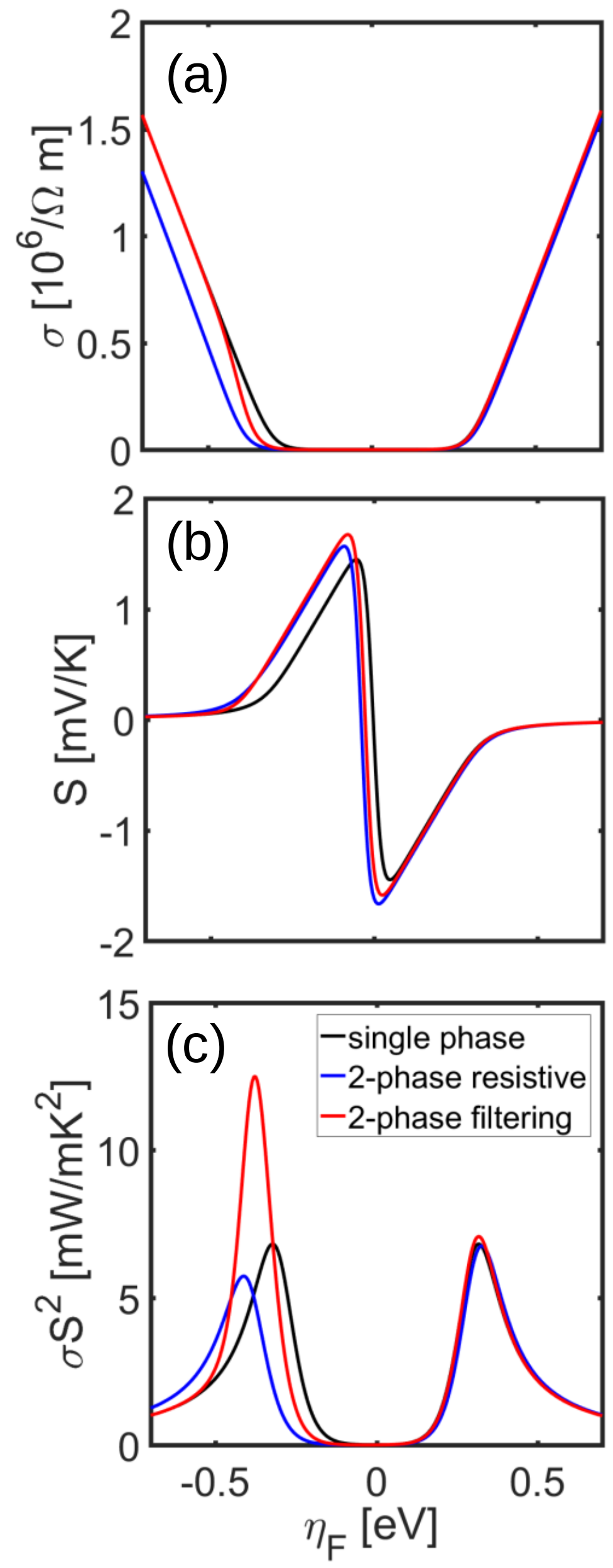

**Fig. 4 Thermoelectric coefficients at 300 K.** (a) Electrical conductivity, (b) Seebeck coefficient, and (c) power factor versus the position of the Fermi level. Three different cases are shown: i) the single phase $Mg_2Si_{0.4}Sn_{0.6}$ system (black lines) which acts as a reference, ii) the two-phase $Mg_2Si_{0.4}Sn_{0.6}$/ $Mg_2Si_{0.6}Sn_{0.4}$ system (blue lines) but under resistive transport conditions (where we assume equal size phases), and iii) the two-phase $Mg_2Si_{0.4}Sn_{0.6}$/$Mg_2Si_{0.6}Sn_{0.4}$ system with the $Mg_2Si_{0.6}Sn_{0.4}$ acting as the filtering barrier (red lines). The mid gap energy of $Mg_2Si_{0.4}Sn_{0.6}$ is set to zero. Elastic acoustic-phonon limited conditions are considered only. Thus, the results intend to illustrate relative performance changes rather than absolute performance levels.